\documentclass[conference]{IEEEtran}
\IEEEoverridecommandlockouts
 \usepackage{xspace}

\usepackage{tikz}
\usepackage{pgfplots}
\pgfplotsset{compat=newest}

\usepackage{tabularx}
\usepackage{graphicx}
\usepackage{balance}

\usepackage{float}
\usepackage{hyperref}
\usepackage{cite}
\usepackage{amsmath,amssymb,amsfonts}
\usepackage{algorithmic}
\usepackage{graphicx}
\usepackage{textcomp}
\usepackage{xcolor}
\def\BibTeX{{\rm B\kern-.05em{\sc i\kern-.025em b}\kern-.08em
    T\kern-.1667em\lower.7ex\hbox{E}\kern-.125emX}}
\begin{document}

\title{GitHub's Copilot Code Review: Can AI Spot Security Flaws Before You Commit?}
\author{\IEEEauthorblockN{Amena Amro, Manar H. Alalfi}
\IEEEauthorblockA{\textit{Department of Computer Science} \\
\textit{Toronto Metropolitan University}
Toronto, ON, Canada\\
{\{amena.amro, manar.alalfi\}@torontomu.ca}}}

\maketitle

\begin{abstract}
As software development practices increasingly adopt AI-powered tools, ensuring that such tools can support secure coding has become critical. This study evaluates the effectiveness of GitHub Copilot’s recently introduced code review feature in detecting security vulnerabilities. Using a curated set of labeled vulnerable code samples drawn from diverse open-source projects spanning multiple programming languages and application domains, we systematically assessed Copilot's ability to identify and provide feedback on common security flaws. Contrary to expectations, our results reveal that Copilot's code review frequently fails to detect critical vulnerabilities such as SQL injection, cross-site scripting (XSS), and insecure deserialization. Instead, its feedback primarily addresses low-severity issues, such as coding style and typographical errors. These findings expose a significant gap between the perceived capabilities of AI-assisted code review and its actual effectiveness in supporting secure development practices. Our results highlight the continued necessity of dedicated security tools and manual code audits to ensure robust software security.
\end{abstract}


\section{Introduction}
In an era where software security is paramount, the integration of AI-powered tools such as GitHub Copilot is transforming the software development process. While Copilot has demonstrated strong capabilities in code generation and developer assistance, its effectiveness in identifying and mitigating security vulnerabilities remains uncertain, particularly as its underlying large language models (LLMs) are continuously evolving~\cite{asleep}. This study investigates the performance of Copilot’s newly introduced code review feature, which, as of February 2025, remains in public preview. Our goal is to assess how well this feature can detect known security flaws in the source code.

To evaluate Copilot’s security review capabilities, we employ labeled vulnerable code samples drawn from a range of publicly available datasets~\cite{datasetAllSafe,dataset2,dataset1,dataset3,dataset4}. These datasets encompass a diverse set of codebases, including web applications, mobile applications, and APIs implemented in multiple programming languages~\cite{dataset2}. One dataset~\cite{dataset1} offers an expanding suite of test cases with documented weaknesses, while another~\cite{dataset3} focuses on known vulnerabilities in widely used libraries and packages. In addition, a vulnerable web service and API~\cite{dataset4} is included to simulate realistic enterprise use cases. All datasets provide ground-truth vulnerability labels, which we use as a benchmark to evaluate Copilot's ability to correctly identify and annotate security issues during code review.

Our evaluation focuses on two key dimensions: (1) the accuracy of Copilot's code review in detecting security vulnerabilities and (2) the clarity and usefulness of its feedback in helping developers understand and resolve these issues. By systematically analyzing Copilot's performance in these areas, we aim to uncover both its strengths and limitations as a security review assistant. Ultimately, our findings contribute to a deeper understanding of the current capabilities of AI-assisted secure coding tools and inform directions for future improvement.

%\paragraph{Research Problem} The research problem addressed in this project is to evaluate the effectiveness of GitHub Copilot's newly introduced code review feature in identifying security vulnerabilities within code. This evaluation will be conducted by comparing Copilot's performance against the CodeQL static analysis engine, focusing on both accuracy in vulnerability detection and the quality of feedback provided for remediation.

%\paragraph{Related Work} ~\cite{asleep} investigated the security implications of the code generated by GitHub Copilot and found that approximately 40\% of the generated programs were found to be vulnerable. The researchers evaluated the generated code using Github’s CodeQL software scanning suite and manual inspection.
%In ~\cite{zeroshot}, the researchers investigated the ability of different LLMs in fixing cybersecurity bugs and how well they are able to provide functional code fixes to historically challenging security bugs. This gives insights into how well LLMs can provide useful feedback in code reviews in the form of functional code fixes to the vulnerable code snippets. 

\section{Background and Related Work}

GitHub Copilot is an AI-powered code completion and synthesis tool developed by GitHub in collaboration with OpenAI. It was originally powered by Codex, a large language model (LLM) trained on a large corpus of public code and natural language. However in 2023 and beyond, GitHub Copilot uses more advanced models such as GPT-4~\cite{chen2021codex}. Originally released to assist developers in writing code by offering inline suggestions, Copilot has recently expanded its functionality to include automated code review. As of early 2025, the Copilot Code Review feature provides natural language feedback on code changes submitted through pull requests. This feature aims to emulate the role of a peer reviewer by identifying potential issues, including security concerns, and suggesting improvements.

Unlike traditional rule-based static analyzers, Copilot’s code review capability is powered by deep learning models trained on vast corpora of publicly available code and associated documentation. Consequently, it does not follow predefined rules or leverage formal program semantics; instead, it relies on learned statistical correlations and contextual patterns~\cite{asleep}. This raises important questions regarding its reliability, particularly in detecting complex or subtle security flaws that may not follow syntactic norms.

Security concerns surrounding LLM-generated code have been well-documented. Perry et al.~\cite{perry2022users} conducted a controlled user study and found that developers using Copilot were more likely to submit insecure solutions compared to those who coded without AI assistance. Alarmingly, these developers were often more confident in their submissions, despite the presence of vulnerabilities. Tihanyi et al.~\cite{tihanyi2024secure} conducted a large-scale empirical study involving over 330,000 C programs generated by multiple LLMs and found that 62\% of the outputs contained at least one security vulnerability. Similarly, Wang et al.~\cite{wang2023codesec} evaluated LLMs on secure code generation tasks and found that while models performed reasonably on code synthesis, they were significantly less effective in repairing scenarios where existing vulnerabilities must be corrected.

The limitations of LLMs in security-focused tasks raise the need to compare them with established static analysis tools. Tools like GitHub’s CodeQL rely on semantic-aware, rule-based detection mechanisms that offer consistent and explainable diagnostics. CodeQL operates by constructing an abstract semantic graph of the code and queries it using a declarative logic-based language to identify common vulnerability patterns. While precise and customizable, CodeQL and similar tools are typically language-specific and may struggle with code written in unconventional styles.

In contrast, LLM-based reviewers such as Copilot are language-agnostic and capable of offering human-like commentary across a broad range of contexts. However, their probabilistic nature leads to inconsistencies and false positives or negatives, especially when facing novel or obfuscated vulnerability patterns~\cite{asleep}. Understanding how Copilot’s code review compares to traditional tools like CodeQL is essential to evaluate its utility for secure software development.

A growing body of related work explores LLMs’ ability not only to generate secure code, but also to assist in code repair and review. Pearce et al.~\cite{asleep} found that approximately 40\% of Copilot-generated programs were vulnerable, even when evaluated against GitHub’s own CodeQL engine. In another relevant study, Pearce et al.~\cite{zeroshot} explored the capabilities of LLMs to fix known security vulnerabilities in a zero-shot setting and found that while some models produced functional patches, success was inconsistent and highly dependent on prompt quality. These studies emphasize the potential value and current limitations of LLMs in supporting secure software engineering workflows.

In light of this evidence, our work focuses specifically on evaluating GitHub Copilot’s Code Review feature: its ability to detect known vulnerabilities in real-world, vulnerable codebases, and the nature of the feedback it provides. To our knowledge, this is one of the first empirical studies evaluating this feature across multiple datasets and application domains.

\section{Methodology}

\subsection{Research Objective}

The primary objective of this study is to empirically evaluate the effectiveness of GitHub Copilot's newly introduced code review feature in identifying security vulnerabilities in source code. Specifically, we evaluate Copilot’s ability to (1) detect known security flaws and (2) provide actionable, contextually appropriate feedback for remediation. As a baseline, we compare Copilot’s performance against GitHub’s CodeQL static analysis engine, focusing on both the detection accuracy and the quality of review comments.

\subsection{Evaluation Design}

To conduct this evaluation, we selected a diverse set of intentionally vulnerable, open-source codebases spanning a range of application domains, including mobile applications, web services, desktop software, and Application Programming Interfaces (APIs). These projects were drawn from authoritative, labeled datasets designed for security research. Each dataset includes source code annotated with ground-truth vulnerability labels, enabling precise comparison between Copilot's review feedback and known security flaws.

\subsection{Benchmark Datasets}

The following datasets were used to evaluate Copilot’s vulnerability detection capabilities:

\begin{itemize}
    \item \textbf{All-Safe}~\cite{datasetAllSafe}: A cross-language dataset of labeled secure and insecure code snippets from multiple domains.
    \item \textbf{VulnCodeDB}~\cite{dataset1}: A curated set of code samples with documented Common Weakness Enumerations (CWEs), intended for benchmarking static analysis tools.
    \item \textbf{Devign}~\cite{dataset3}: A dataset containing real-world security bugs extracted from open-source GitHub repositories.
    \item \textbf{WebGoat API}~\cite{dataset4}: A deliberately insecure RESTful API developed by Open Worldwide Application Security Project (OWASP) for secure coding education and evaluation.
    \item \textbf{VulApps}~\cite{dataset2}: A suite of vulnerable applications spanning web, mobile, and service-oriented architectures.
\end{itemize}

These datasets were chosen for their diversity, realistic vulnerability scenarios, and comprehensive ground-truth labeling, making them suitable for assessing both vulnerability detection and review effectiveness.

\subsection{Experimental Procedure}

For each selected project, we performed the following steps:

\begin{enumerate}
    \item \textbf{Repository Setup}: Each codebase was hosted in a GitHub repository that was set to trigger automatic reviews from Copilot for each pull request (PR) against the main branch. Vulnerable code segments were introduced through pull requests (PRs), simulating a standard developer workflow.
    
    \item \textbf{Triggering Copilot Review}: PRs were submitted to invoke GitHub Copilot's code review.
    
    \item \textbf{Data Collection}: We recorded Copilot’s review output for each PR, including:
    \begin{itemize}
        \item The number of files reviewed.
        \item The number and types of comments generated (e.g., suggestions, typo corrections, security observations).
        \item Specific security flaws identified or referenced, either directly (e.g., “This may be an SQL injection”) or indirectly (e.g., generic warnings about input validation).
    \end{itemize}
    
    \item \textbf{Ground-Truth Comparison}: We compared Copilot’s comments with the known vulnerabilities annotated in the dataset to determine whether the tool successfully identified each flaw, missed it, or generated false positives.
    
    \item \textbf{Re-review Analysis}: For a subset of examples, we repeated the review process by resubmitting semantically identical code with minor syntactic changes to evaluate Copilot’s consistency across different iterations.
\end{enumerate}

All evaluations were conducted using the automatic GitHub Copilot code review tool without prompt engineering (except for the re-review analysis where a direct prompt of "review code" was used), external configuration, or manual tuning. This ensures ecological validity by approximating the experience of a typical developer using Copilot in a production-like setting.

\subsection{Language Support Constraints}

At the time of writing, GitHub Copilot’s Code Review feature supports a fixed set of programming languages, shown in Table~\ref{tab:table1}. All selected datasets were filtered to include files written in supported languages to ensure compatibility.

\begin{table}[t]
    \centering
    \begin{tabular}{llllll}
        C & C\# & C++ & Go & Java & JavaScript \\
        Kotlin & Markdown & Python & Ruby & Swift & TypeScript \\
    \end{tabular}
    \caption{Languages Supported by GitHub Copilot Code Review}
    \label{tab:table1}
\end{table}

This methodology provides a controlled and reproducible framework for evaluating the security analysis capabilities of Github's Copilot code review tool across multiple programming languages and application domains.

\section{Experiment results}
In this section, we present the evaluation results of GitHub Copilot's ability to detect security flaws during code reviews. We analyzed multiple datasets across different application domains and programming languages.

\section{Discussion}

This section analyzes the results summarized in Table~\ref{tab:evaluation}, focusing on GitHub Copilot's performance in reviewing security-relevant code across a diverse set of domains, languages, and vulnerability types. We examine not only Copilot's success and failure in vulnerability detection, but also its feedback quality, consistency, and practical applicability in secure software engineering workflows.

\subsection{General Observations Across Datasets}

Overall, GitHub Copilot's Code Review feature demonstrated a low success rate in identifying known security vulnerabilities in purposefully vulnerable codebases. Despite reviewing the vast majority of files across projects, Copilot’s output was often minimal or entirely unrelated to security. Across 7 benchmark datasets, which collectively included hundreds of documented vulnerabilities (e.g., SQL injection, insecure deserialization, cross-site scripting), Copilot generated a total of fewer than 20 comments, most of which addressed spelling or minor style issues.

This is particularly concerning given that many of these datasets such as SARD and WebGoat are widely used in both academia and industry to benchmark the detection capabilities of security tools. The failure to detect even one instance of a critical vulnerability (e.g., SQL injection or XSS) strongly indicates that Copilot’s current review model is not security-aware in any practical sense.

\subsection{Dataset-Specific Insights}

\subsubsection{Allsafe (Mobile Apps: Java/Kotlin/C++)}
\label{appendix:allsafe}

\begin{figure}[h!]
    \centering
    \includegraphics[width=\linewidth]{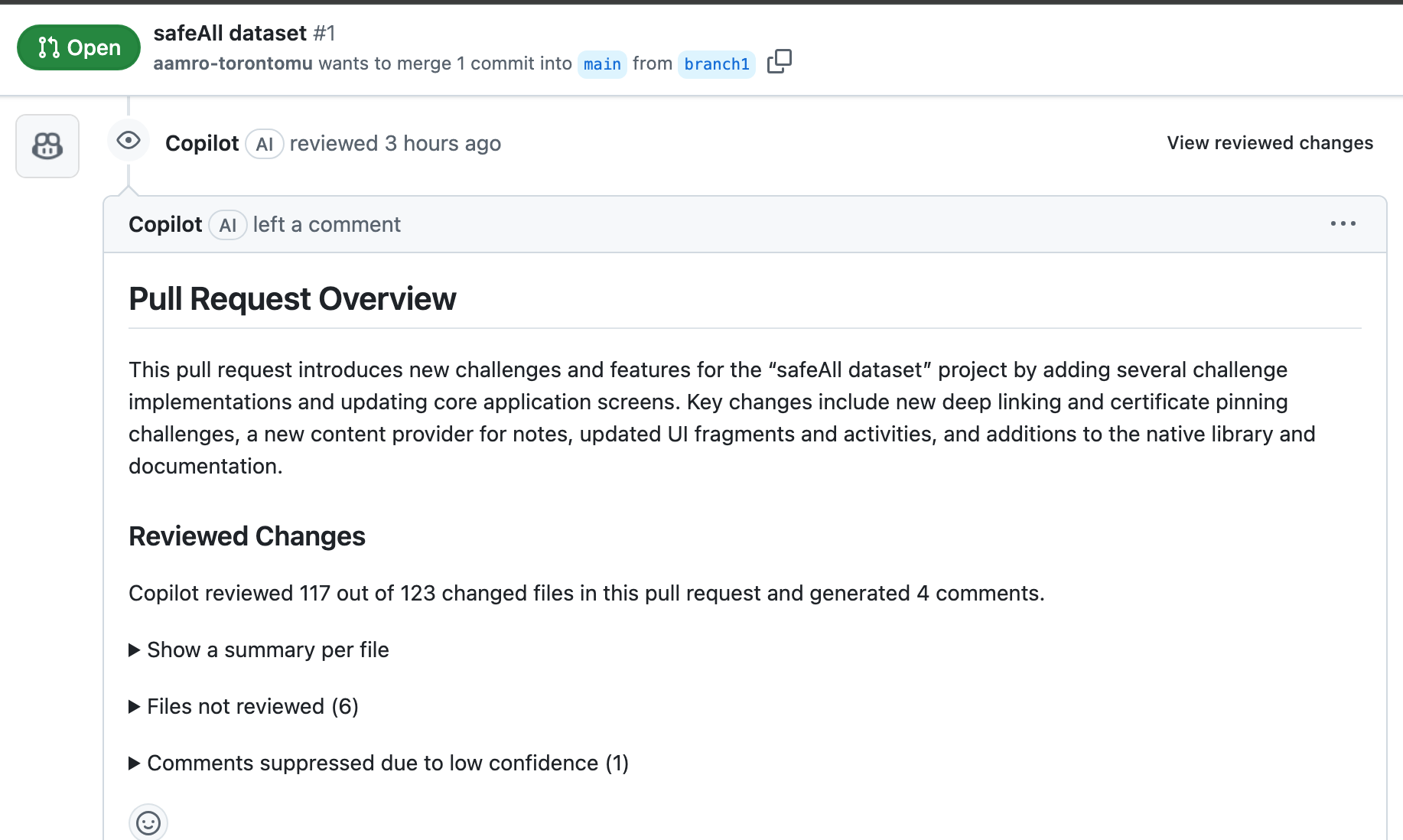} 
    \caption{Allsafe Dataset Details}
    \label{fig:allsafeDetail}
\end{figure}

\begin{figure}[h!]
    \centering
    \includegraphics[width=\linewidth]{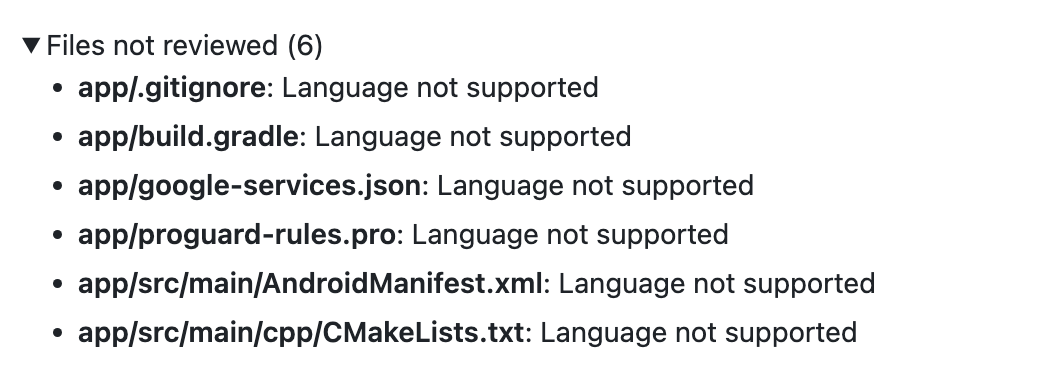} 
    \caption{Allsafe list of not reviewed files}
        \label{fig:allsafeDetail1}
\end{figure}

\begin{figure}[h!]
    \centering
    \includegraphics[width=\linewidth]{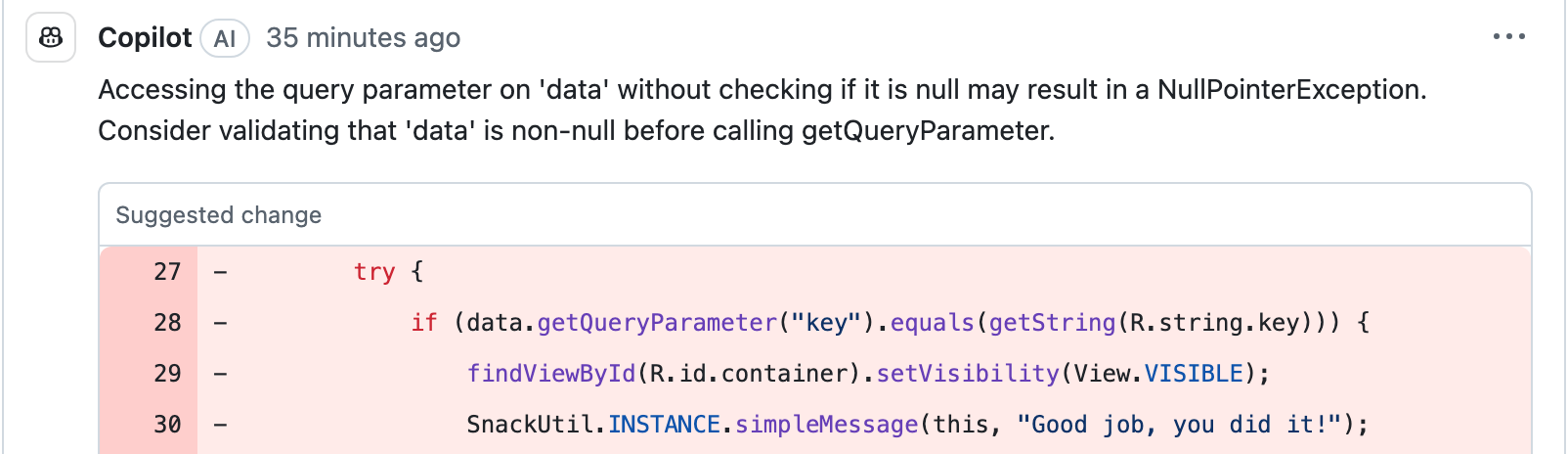} 
    \caption{Allsafe Comment 1: NullPointerException (with suggested changes).}
    \label{fig:allsafeDetail2}
    \end{figure}

\begin{figure}[h!]
    \centering
    \includegraphics[width=\linewidth]{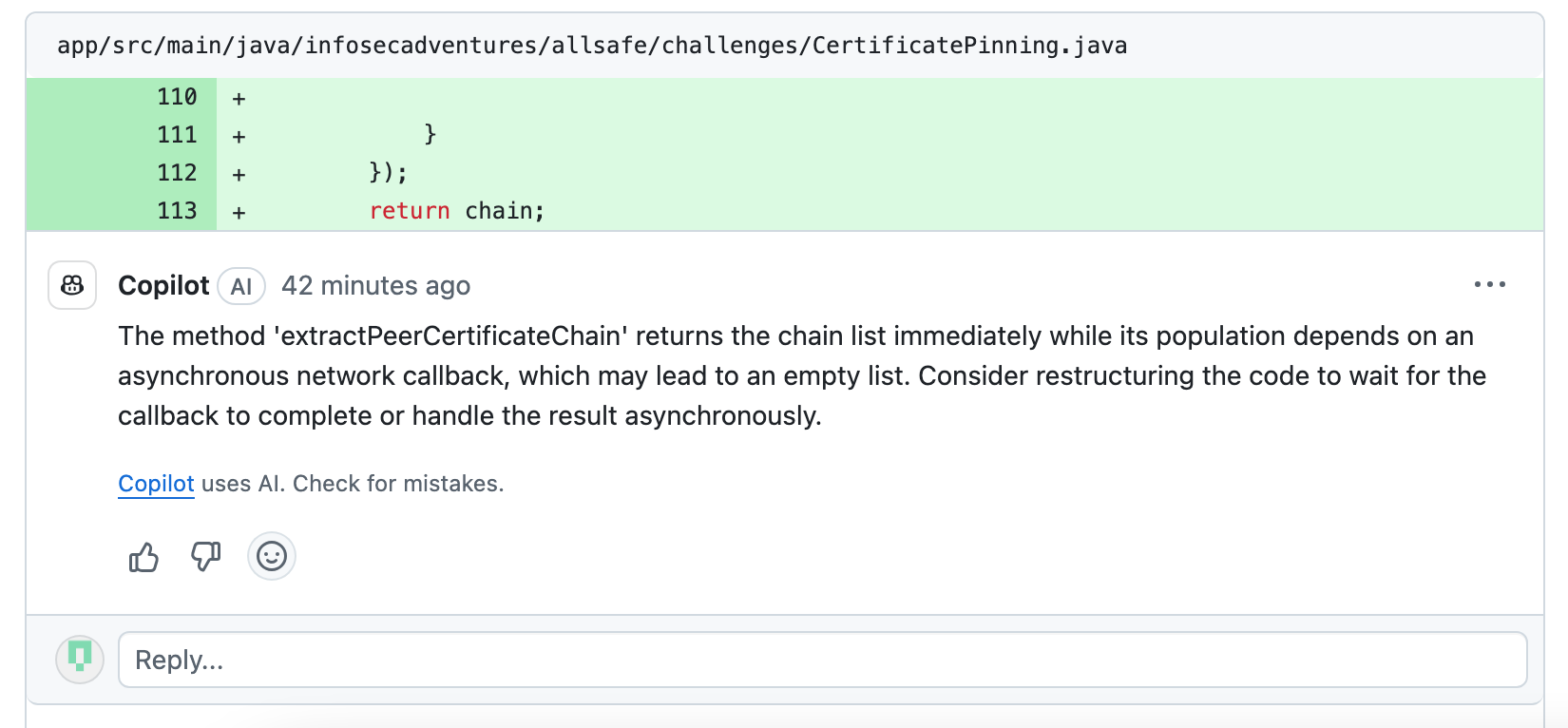} 
    \caption{Allsafe Comment 2 and 3: irrelevant (to security).}
    \label{fig:allsafeDetail13}
\end{figure}

\begin{figure}[h!]
    \centering
    \includegraphics[width=\linewidth]{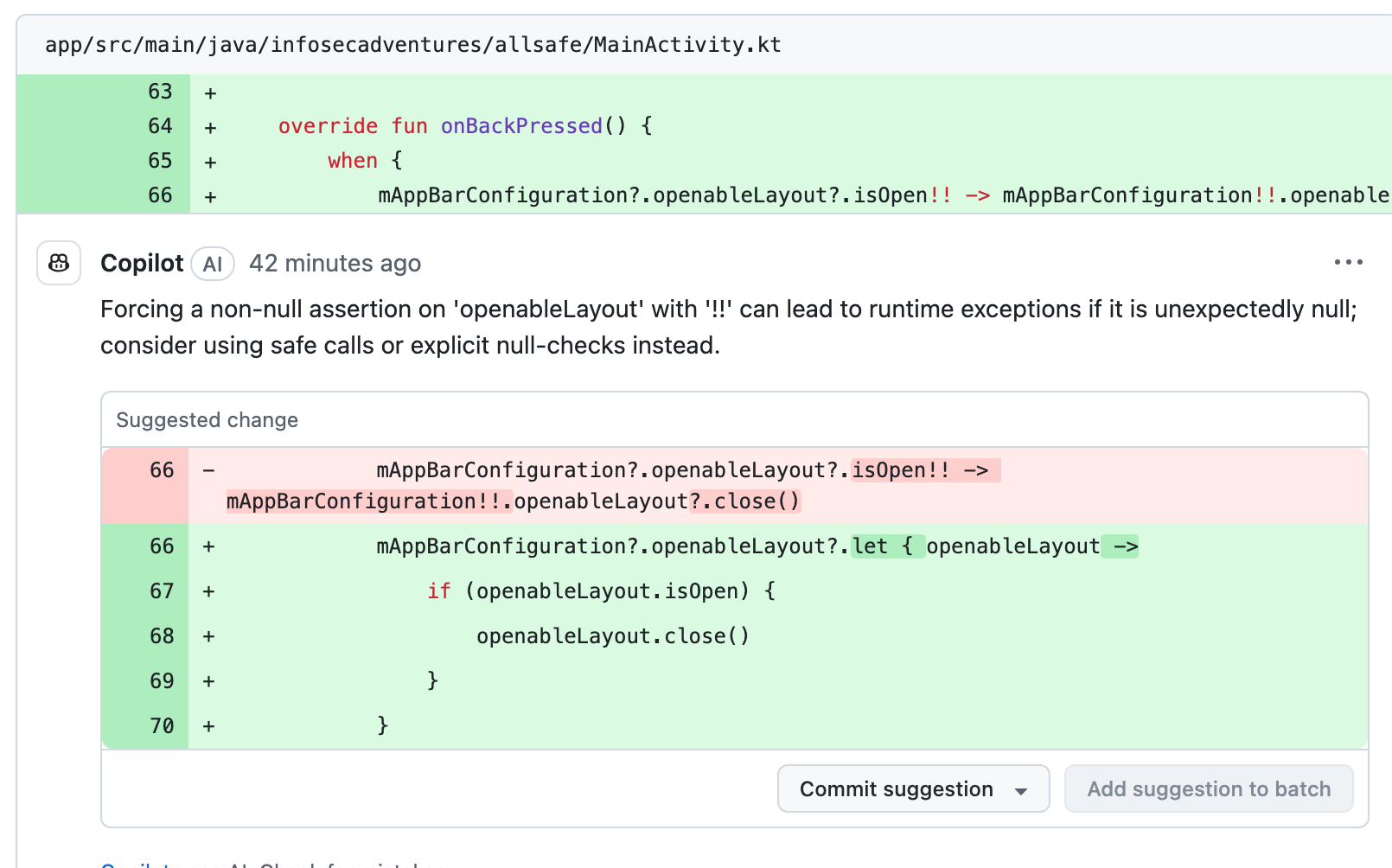} 
    \caption{Allsafe Comment 4: runtime exception (with suggested changes)}
    \label{fig:allsafeDetail4}
\end{figure}

The \textbf{Allsafe} dataset consists of intentionally insecure mobile applications with rich, varied vulnerabilities, including insecure logging, deep link exploitation, and certificate pinning bypass. Despite reviewing 117 out of 123 files, Copilot produced only 4 comments as shown in Figure~\ref{fig:allsafeDetail}, none of which referenced any vulnerability. Notably, \texttt{AndroidManifest.xml}—a crucial file for detecting permission misuse and insecure component exposure—was not reviewed at all. Details are in Figure~\ref{fig:allsafeDetail1}, Figure~\ref{fig:allsafeDetail2}, Figure~\ref{fig:allsafeDetail13}, and Figure~\ref{fig:allsafeDetail4}. This highlights a significant blind spot in Copilot’s coverage: platform-specific configuration files, which are often the source of mobile security issues, are ignored entirely. While Copilot Code Review feature was not able to review AndroidManifest.xml file as shown in the image below, \textbf{Ask Copilot} feature is able to analyze and detect vulnerabilities in the same file when prompted to code-review it. 

%\subsubsubsection{Appendix A: Allsafe Dataset Details}

%Copilot reviewed 117 out of 123 changed files in this pull request and generated 4 comments \ref{fig:allsafeDetail}.

\subsubsection{WebGoat.NET (Web App: C\# / ASP / JavaScript)}
\label{appendix:webgoat}
\begin{figure}[h!]
    \centering
    \includegraphics[width=\linewidth]{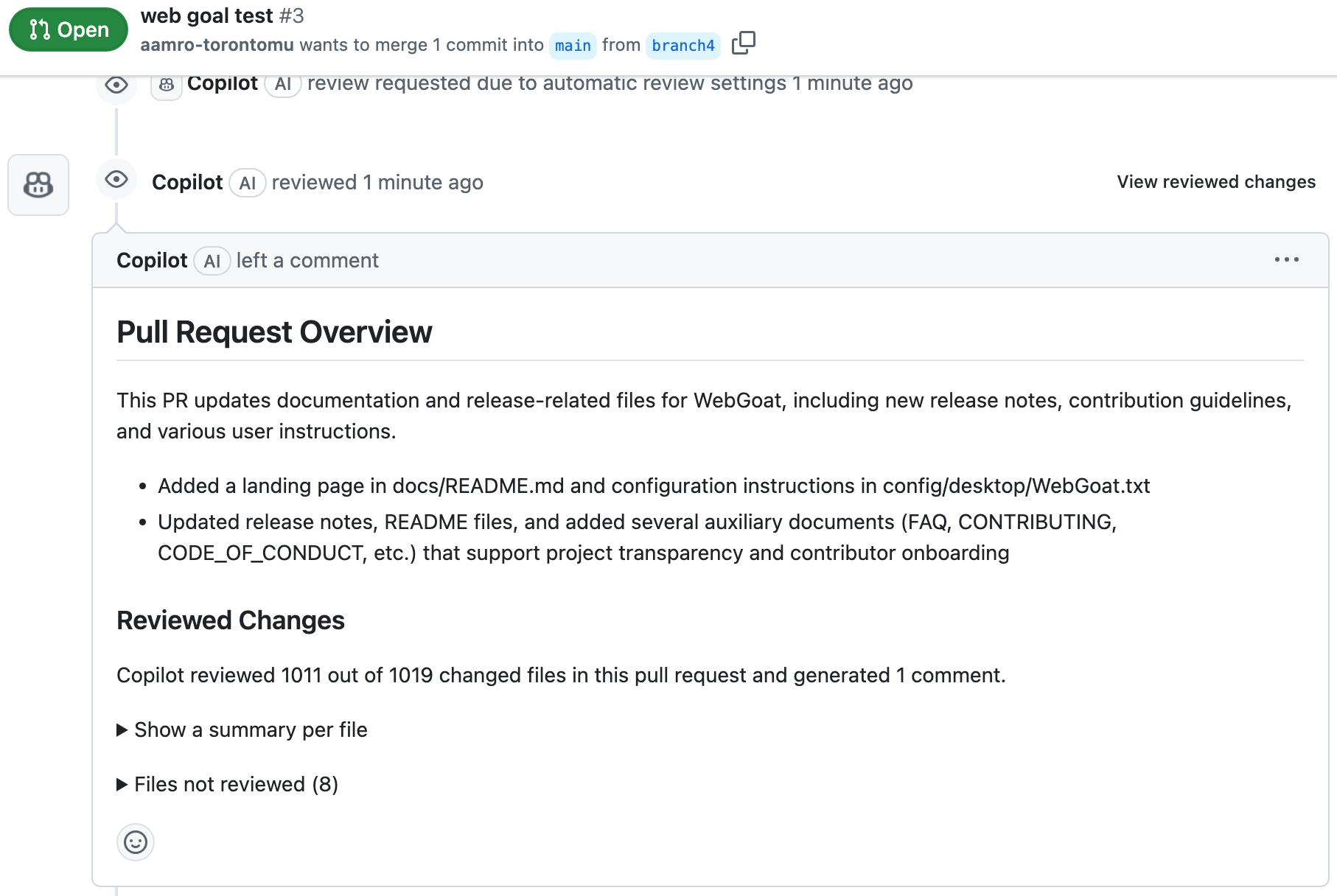} 
    \caption{WebGoat Dataset Details}
    \label{fig:webgoat1}
\end{figure}

\begin{figure}[h!]
    \centering
    \includegraphics[width=\linewidth]{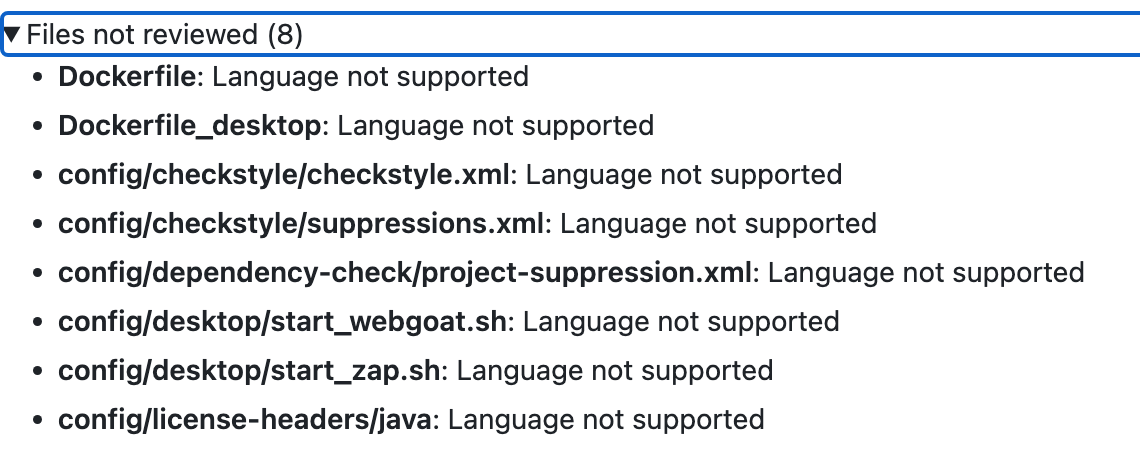} 
    \caption{WebGoat Dataset-Files not reviewed}
    \label{fig:webgoat2}
\end{figure}

\begin{figure}[h!]
    \centering
    \includegraphics[width=\linewidth]{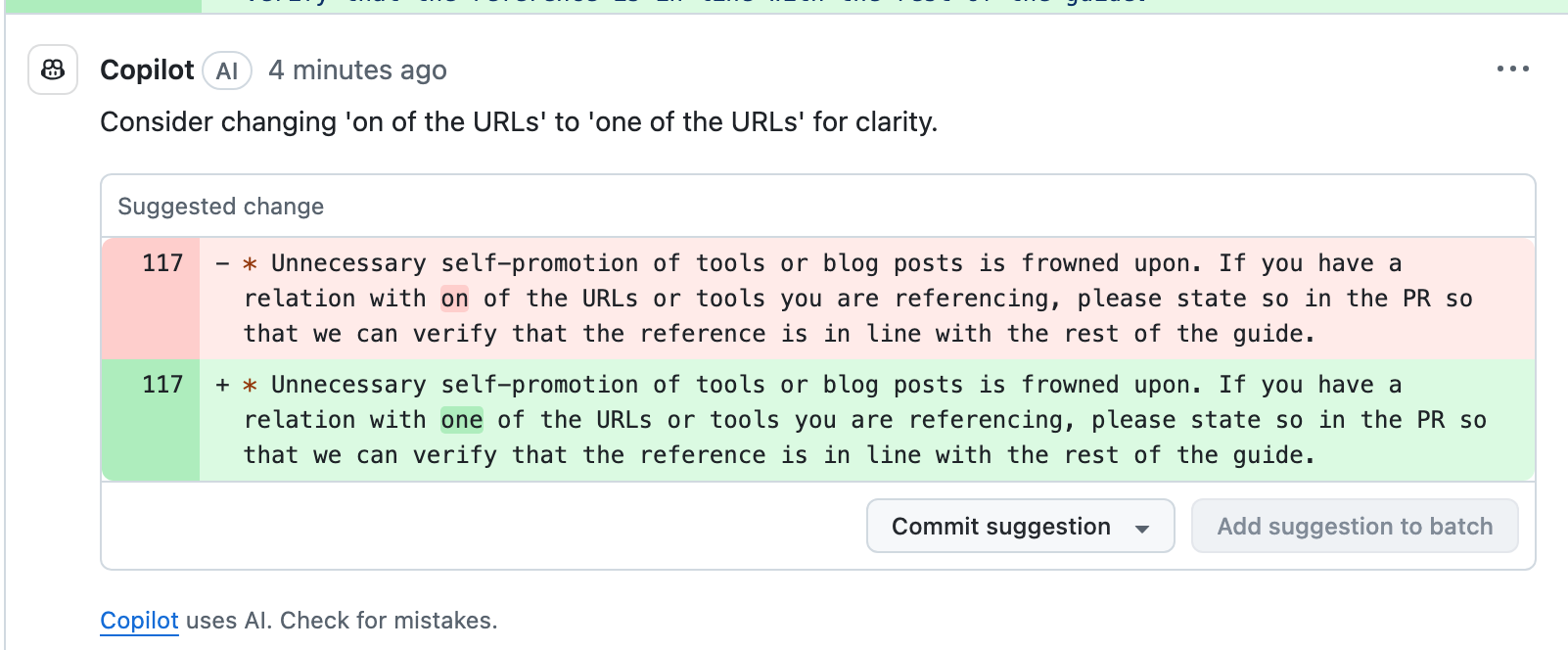} 
    \caption{WebGoat Comment 1: spelling check of a comment.}
    \label{fig:webgoat3}
\end{figure}

\begin{figure}[h!]
    \centering
    \includegraphics[width=\linewidth]{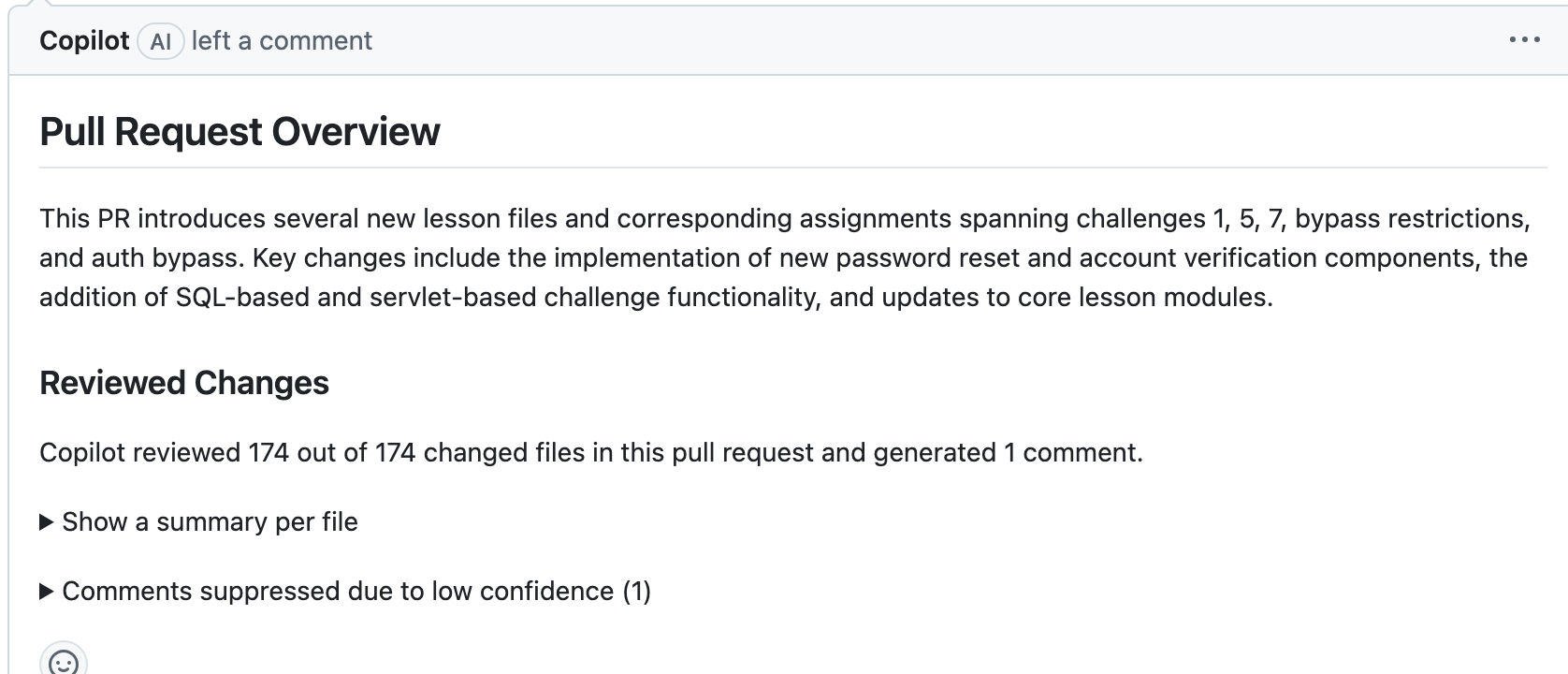} 
    \caption{WebGoat Selected Java files.}
    \label{fig:webgoat4}
\end{figure}

\begin{figure}[h!]
    \centering
    \includegraphics[width=\linewidth]{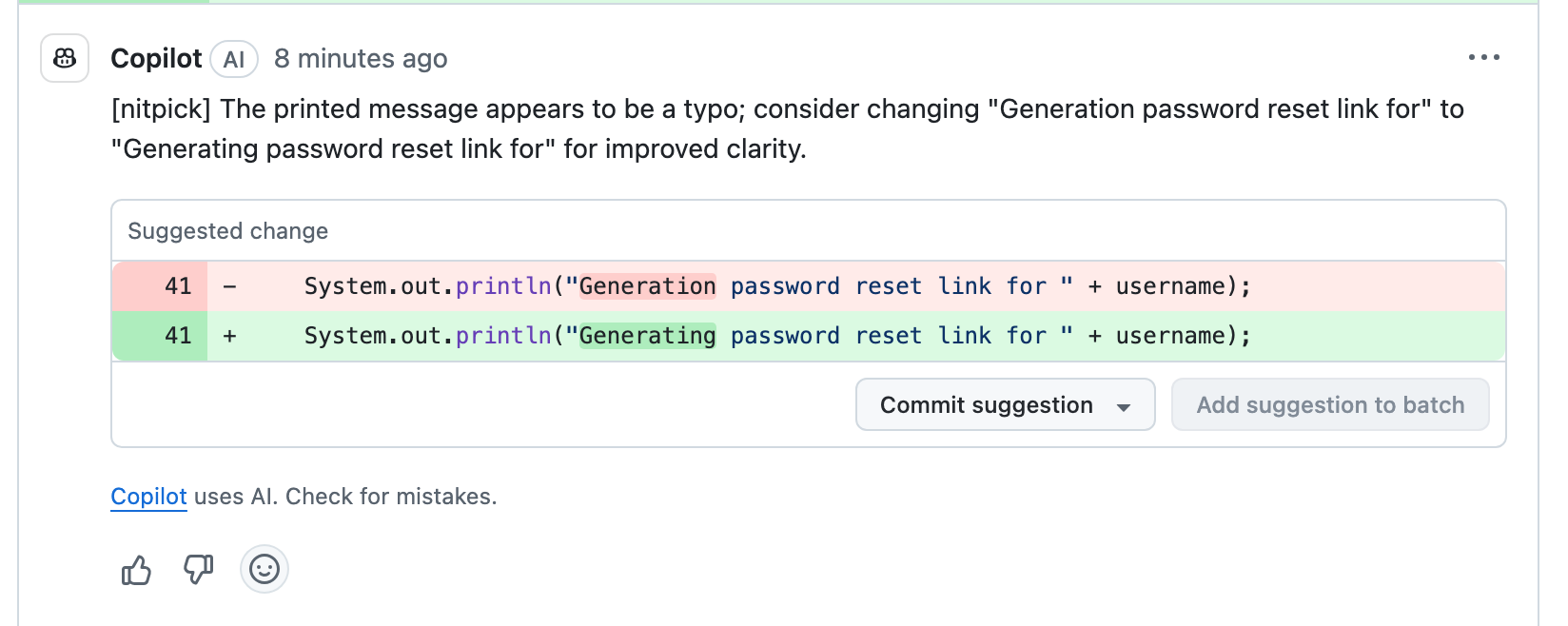} 
    \caption{WebGoat Comment 1: another spelling check of a comment}
    \label{fig:webgoat5}
\end{figure}

\textbf{WebGoat} is a deliberately vulnerable project used to train developers on secure coding practices. It includes classic web security flaws such as broken authentication, XSS, and insecure deserialization. Figures~\ref{fig:webgoat1} and ~\ref{fig:webgoat2} shows that Copilot reviewed 1011 out of 1019 files and generated only a single comment about a typographical error, Figure~\ref{fig:webgoat3}. Even when narrowed to Java files (174/174) as in Figure~\ref{fig:webgoat4}, feedback remained trivial (Figure~\ref{fig:webgoat5}). This is particularly troubling, as WebGoat is explicitly designed to contain known OWASP Top 10 vulnerabilities—yet none were flagged.

\begin{figure}[h!]
    \centering
    \includegraphics[width=\linewidth]{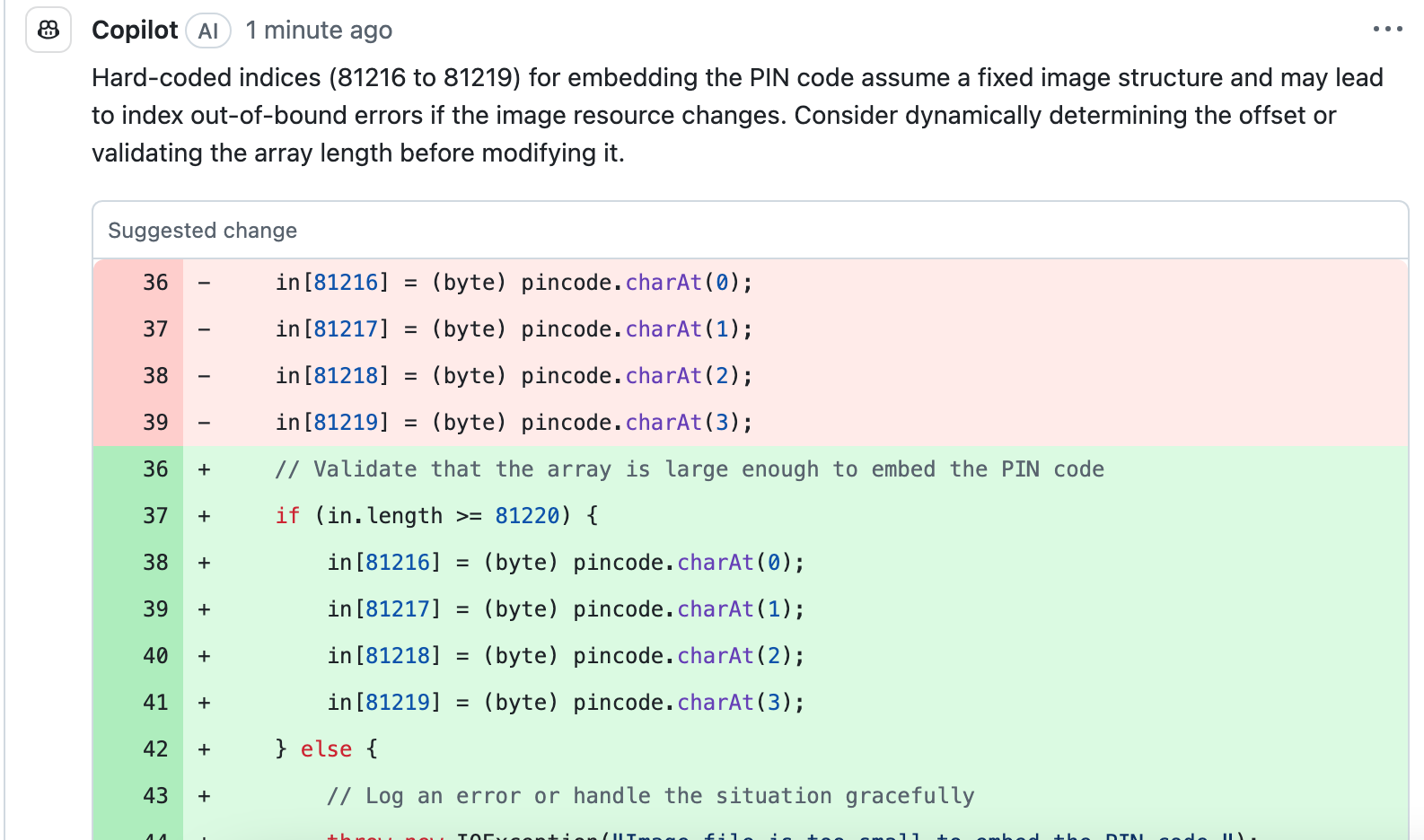} 
    \caption{WebGoat On Re-Review Request, one comment: hardcoded indices}
    \label{fig:webgoat6}
\end{figure}

Upon re-review (Figure~\ref{fig:webgoat6}), Copilot identified a hardcoded index in an array, which, while relevant to robustness, was unrelated to security. This case illustrates that Copilot’s detection may occasionally stumble upon bad practices, but lacks a coherent model for identifying security risks unless they coincide with more obvious code smells.
%\subsection{Appendix B: WebGoat Dataset Details}
%\label{appendix:webgoat}
Entire Codebase. Copilot reviewed 1011 out of 1019 changed files in this pull request and generated 1 comment.

%Comment 1: spelling check of a comment.

%Selected Java files.
%Comment 1: another spelling check of a comment
%On Re-Review Request, one comment: hardcoded indices
\subsubsection{SARD: Test Cases (XSS and SQL Injection)}
\label{appendix:cwe79}
\begin{figure}[h!]
    \centering
    \includegraphics[width=\linewidth]{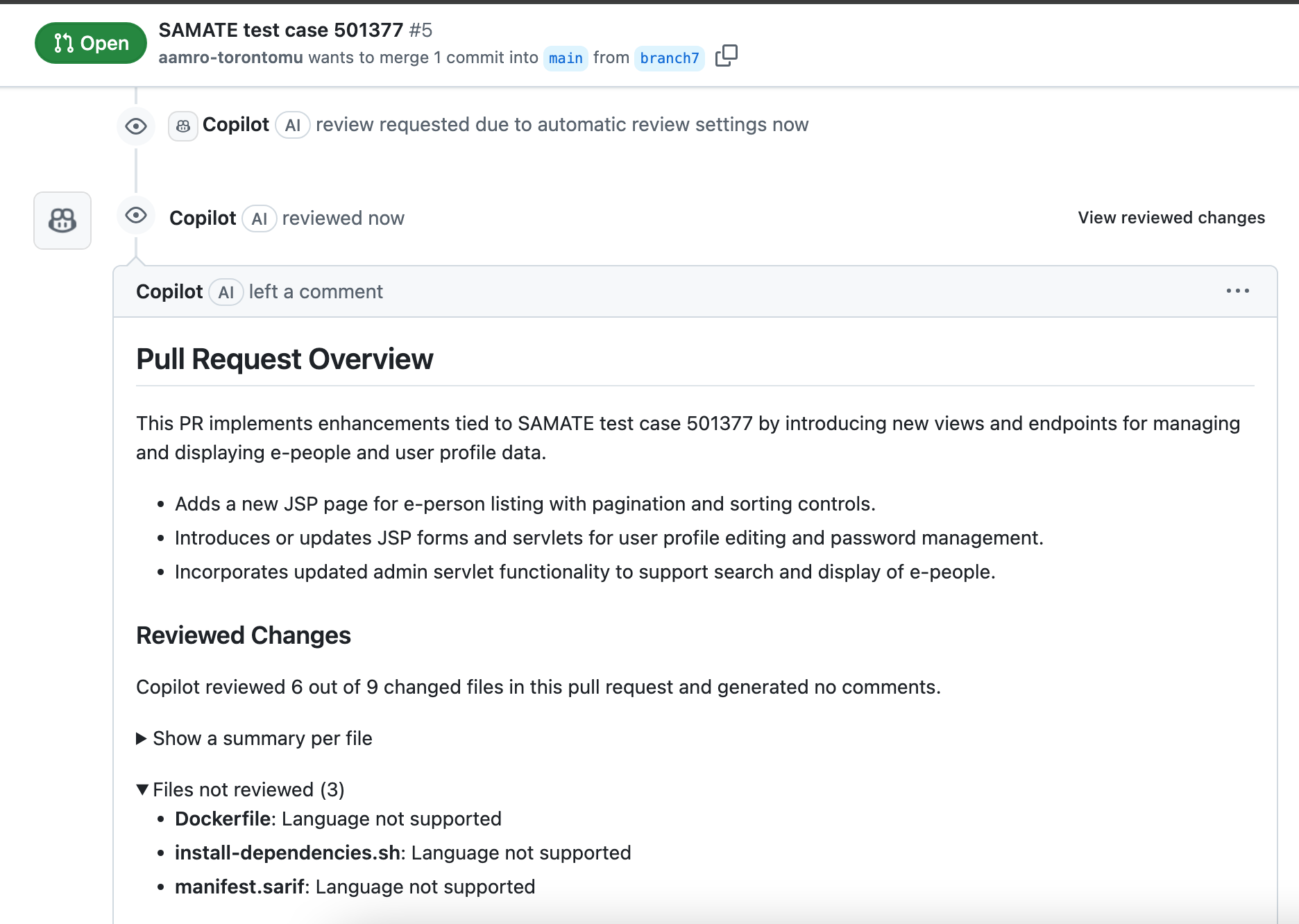} 
    \caption{SARD XSS Test Case Details}
    \label{fig:sard1}
\end{figure}

\begin{figure}[h!]
    \centering
    \includegraphics[width=\linewidth]{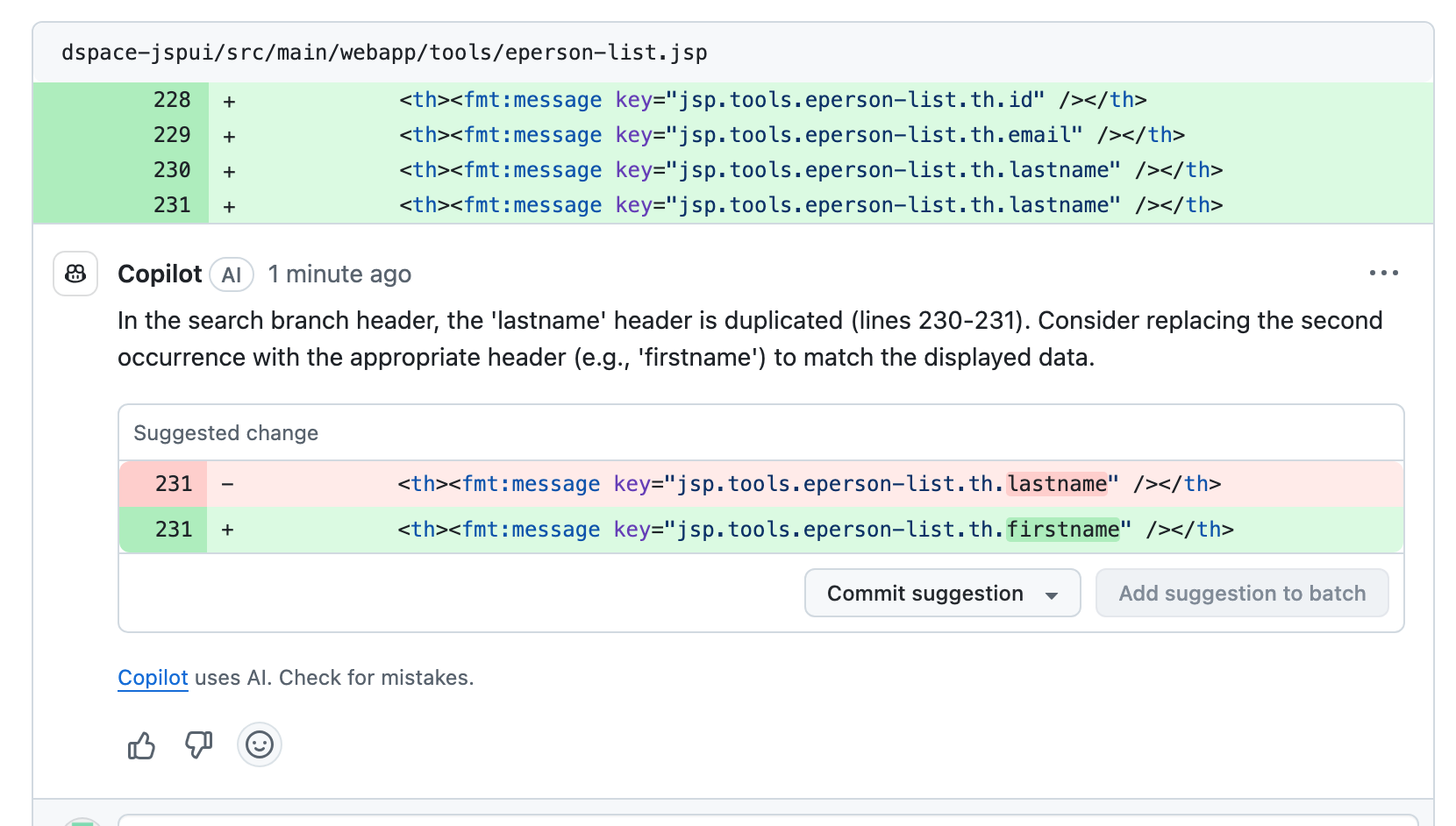} 
    \caption{SARD On re-review request}
    \label{fig:sard2}
\end{figure}

SARD test cases offer isolated, synthetic examples of specific CWEs (e.g., CWE-79 XSS and CWE-564 SQL injection). These test cases are precisely labeled and relatively short—ideal conditions for a reviewer. Yet, for the XSS case, Copilot reviewed 6 of 9 files and made no comments. For the SQL injection case, it reviewed 4 of 9 files and flagged only spelling mistakes (Figures ~\ref{fig:sard1} and ~\ref{fig:sard2}).

This suggests Copilot either failed to interpret the execution flow leading to injection or lacked the semantic reasoning to associate string concatenation in SQL queries with exploitable paths—something traditional static analyzers routinely catch.
%\subsubsubsection{Appendix C: SARD XSS Test Case Details}

%On re-review request
\subsubsection{IARPA STONESOUP / Wireshark (Desktop App: C)}
\label{appendix:wireshark}
\begin{figure}[h!]
    \centering
    \includegraphics[width=\linewidth]{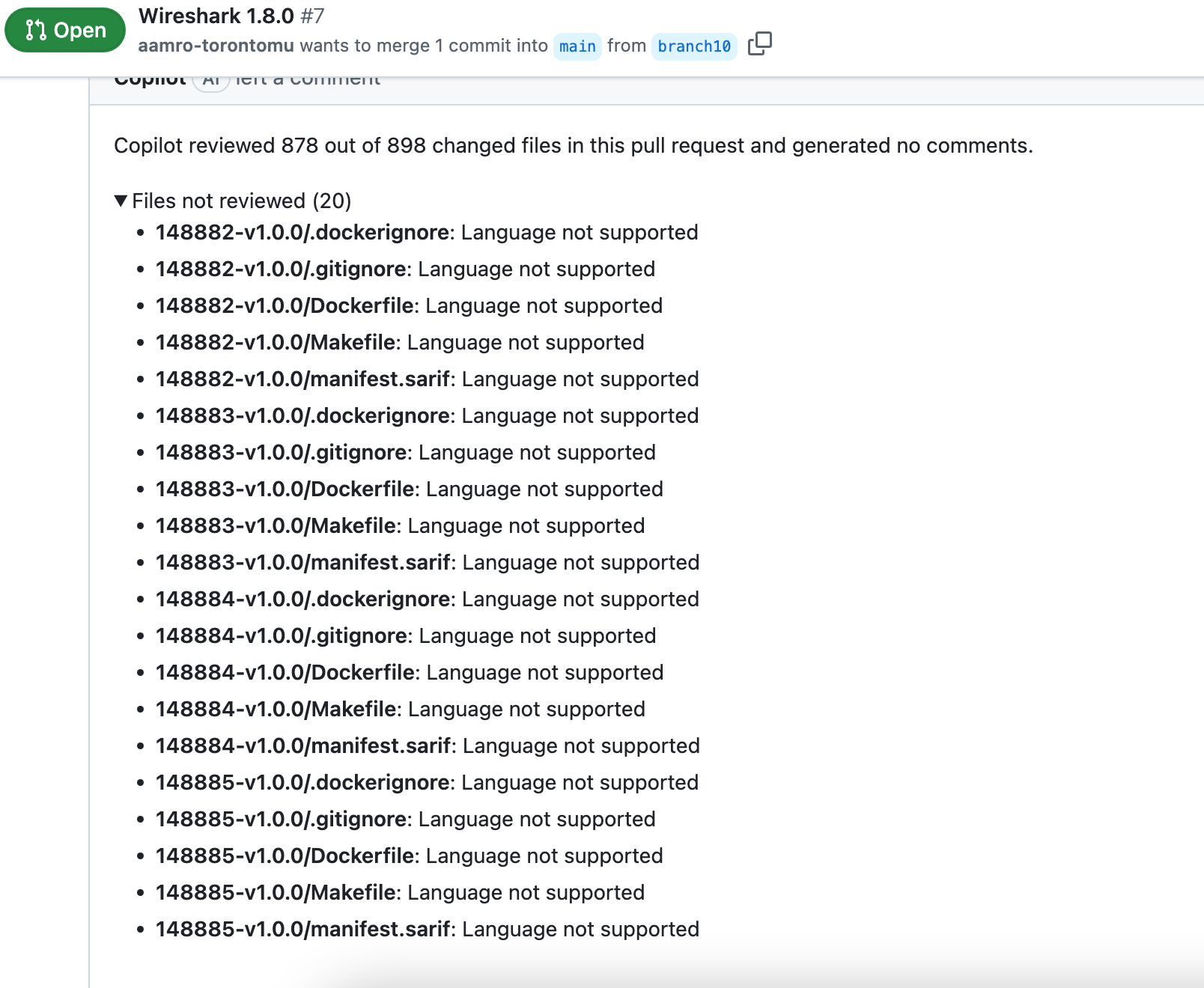} 
    \caption{Wireshark: Copilot reviewed 878 out of 898 changed files in this pull request and generated no comments.}
    \label{fig:wireshark}
\end{figure}

The Wireshark test suite from the STONESOUP program includes a large number of real-world vulnerabilities across multiple CWEs, including buffer overflows, memory allocation issues, and null termination bugs. Copilot reviewed 878 out of 898 files with zero comments (Figure ~\ref{fig:wireshark}). This is particularly alarming given that memory corruption bugs in C remain among the most critical and historically exploited vulnerabilities.

The likely cause of this failure is twofold: first, Copilot may lack effective training data on low-level C programming idioms; second, it lacks the deep semantic analysis required to track pointer arithmetic, bounds, and control flow across functions—tasks better suited to symbolic execution and static analysis tools like CodeQL or Frama-C.
%\subsubsubsection{Appendix D: Wireshark 1.8.0 Test suite Details}
%\label{appendix:wireshark}
IARPA STONESOUP phase 3 - Wireshark 1.8.0 Test suite 22.
Copilot reviewed 878 out of 898 changed files in this pull request and generated no comments.

\subsubsection{SARD Sensitive Cookie (Web App: Java)}
\label{appendix:cookie}
\begin{figure}[h!]
    \centering
    \includegraphics[width=\linewidth]{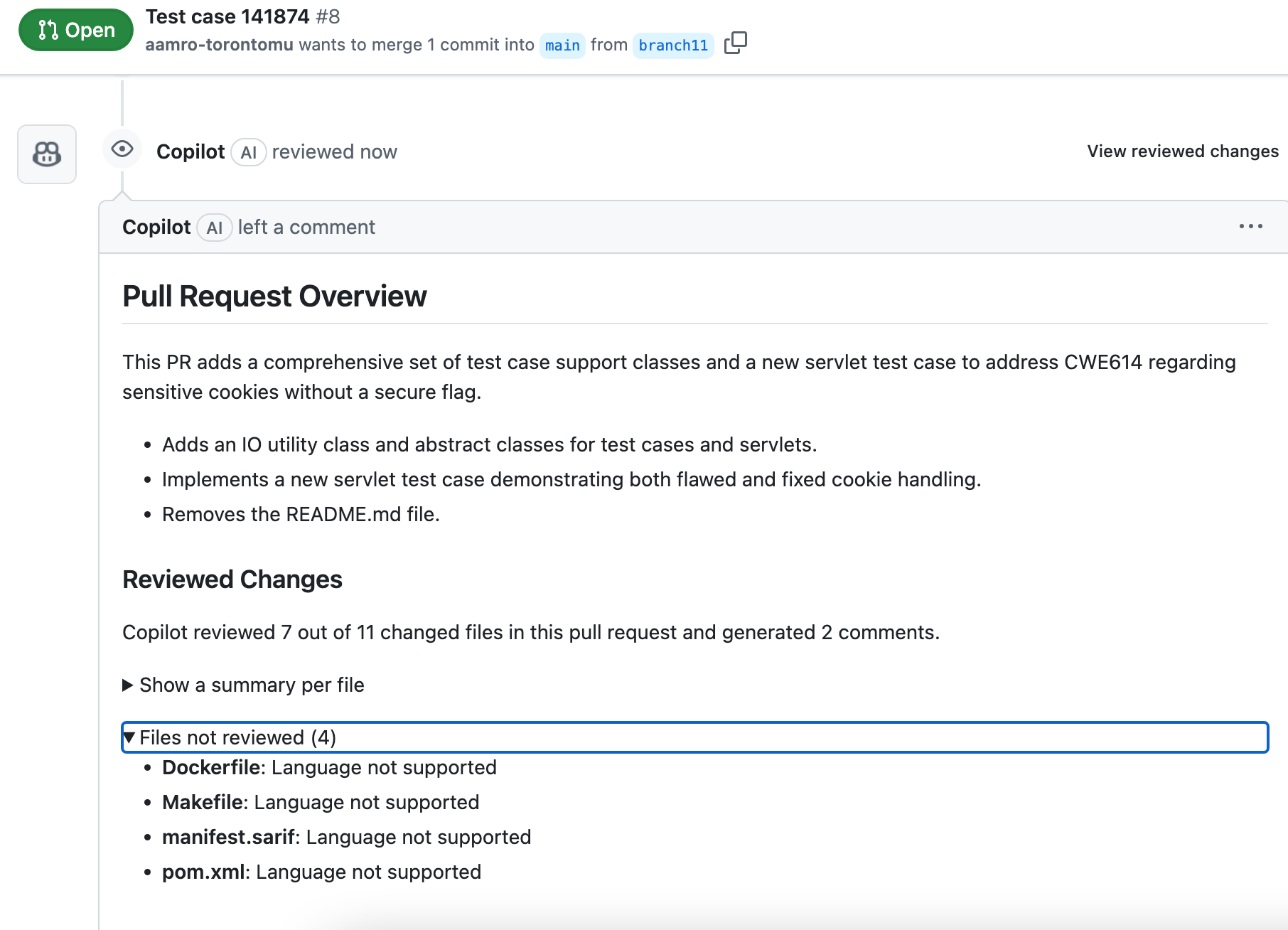} 
    \caption{SARD Sensitive Cookie: Both Comments are spelling mistakes.}
    \label{fig:SARDcookie}
\end{figure}

This dataset includes a CWE-614 vulnerability (missing secure flag on a cookie). Copilot reviewed 7 out of 11 files and produced two comments (Figure ~\ref{fig:SARDcookie}), again related to spelling. The failure to detect a common web security misconfiguration, which can be statically identified by inspecting HTTP header generation, illustrates that Copilot does not generalize across vulnerability types—even those that appear in explicit, declarative code constructs.
%\subsubsubsection{Appendix E: Sensitive
%Cookie Test Details}
%\label{appendix:cookie}
%Both Comments are spelling mistakes.

\subsubsection{dvws-node (Web API: JavaScript)}
\label{appendix:dvws}
\begin{figure}[h!]
    \centering
    \includegraphics[width=\linewidth]{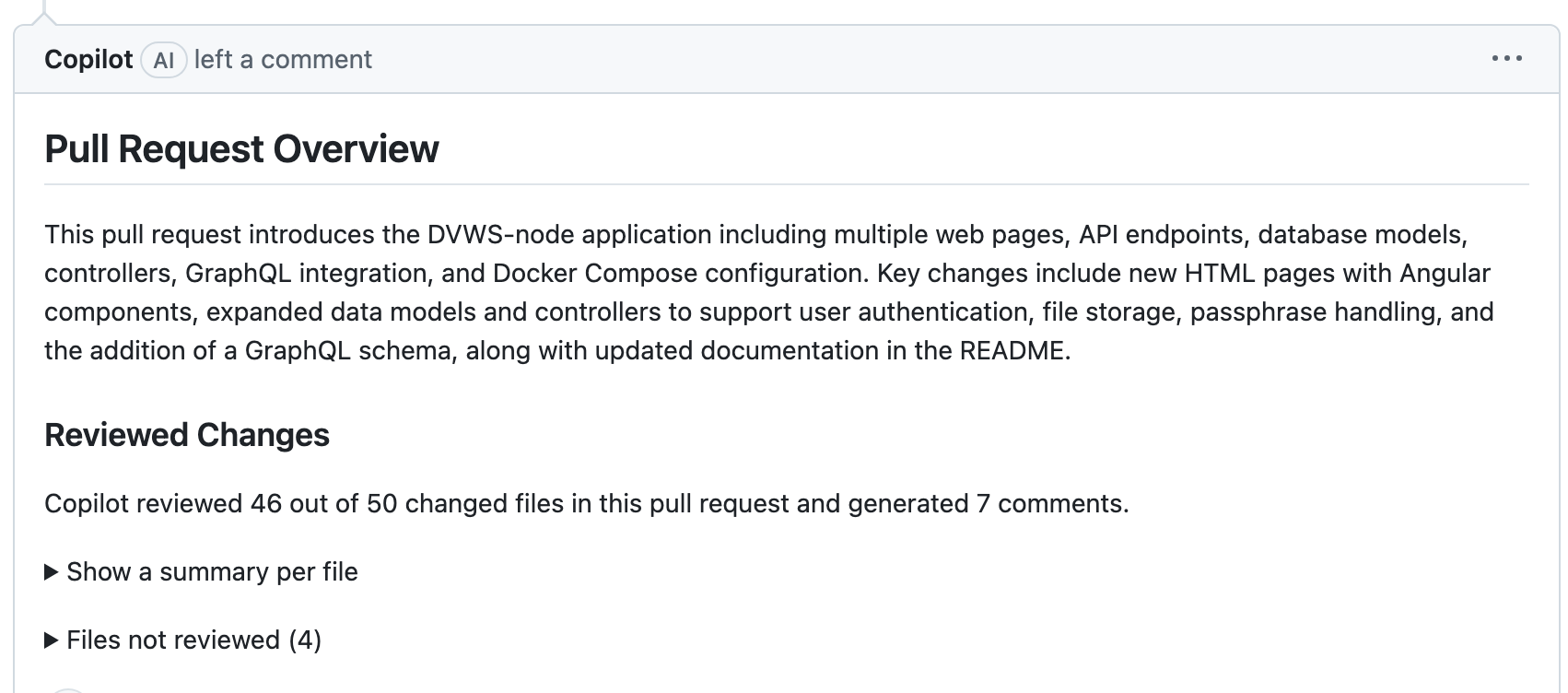} 
    \caption{dvws-node: On Re-Review Request, Copilot reviewed 46 out of 50 changed files in this pull request and generated 7 comments.}
    \label{fig:dvws1}
\end{figure}

\begin{figure}[h!]
    \centering
    \includegraphics[width=\linewidth]{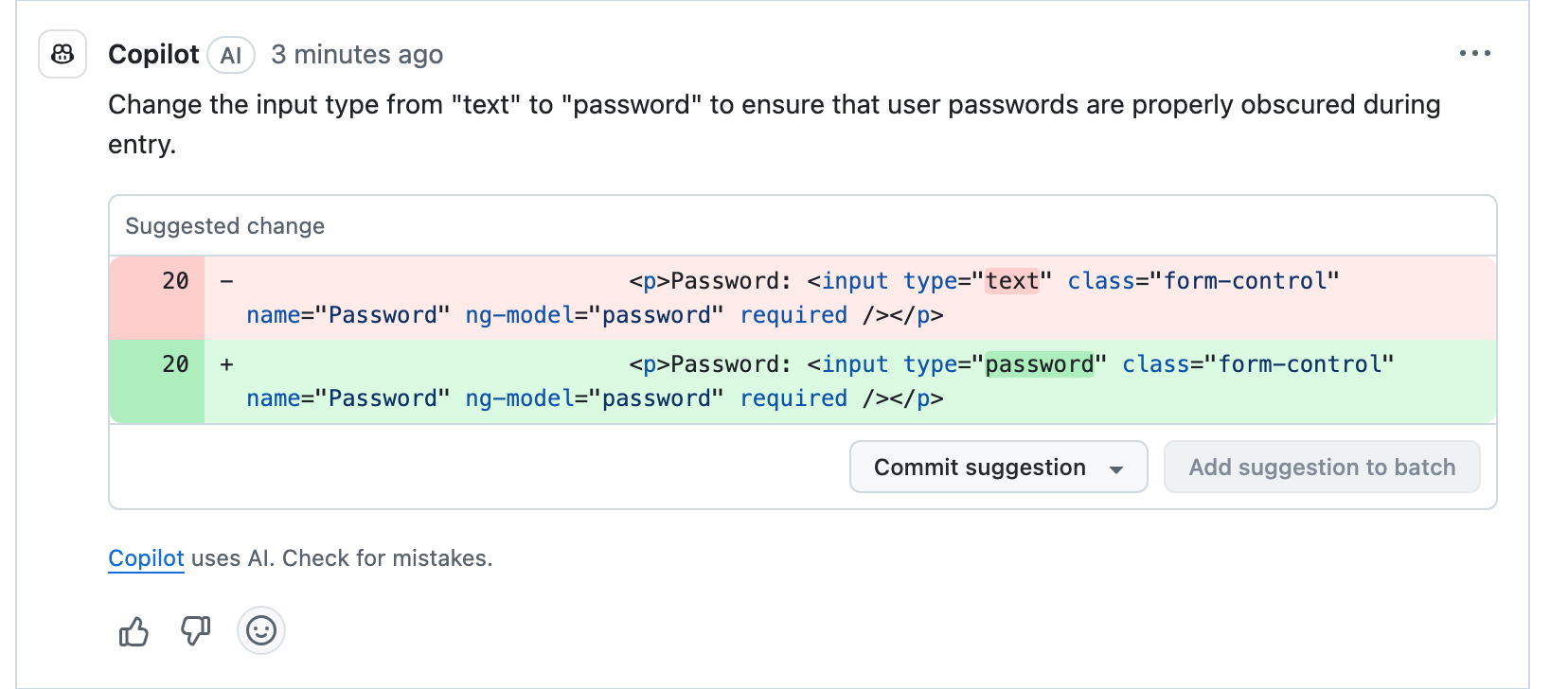} 
    \caption{dvws-node: properly identify password field}
    \label{fig:dvws2}
\end{figure}

\begin{figure}[h!]
    \centering
    \includegraphics[width=\linewidth]{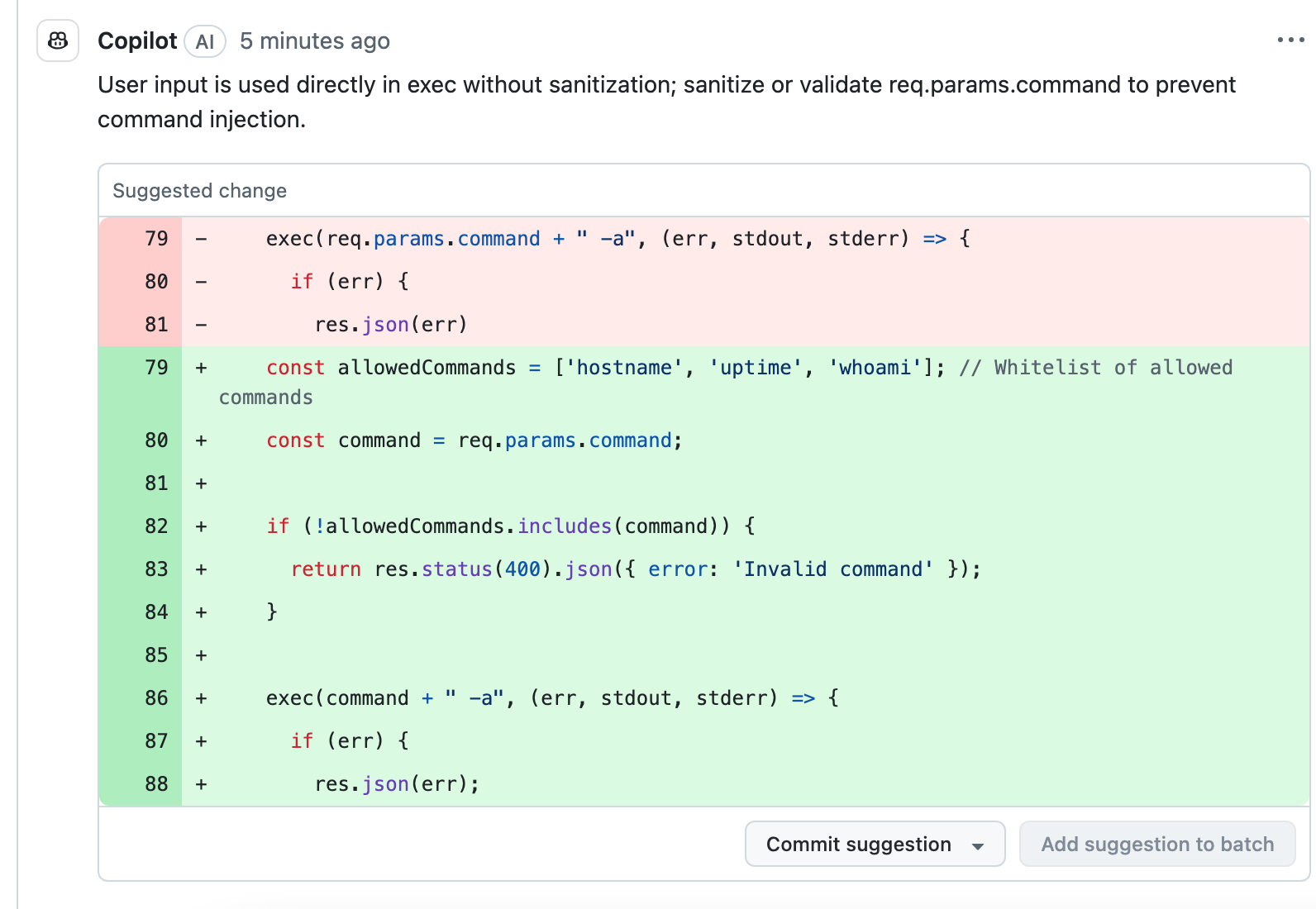} 
    \caption{dvws-node: improper sanitization, spelling mistake}
    \label{fig:dvws3}
\end{figure}

This JavaScript-based project includes a variety of injection and access control vulnerabilities. In its initial run, Copilot reviewed 46 out of 50 files and produced no comments. Upon re-review, 7 comments were generated, some of which touched on password input fields and sanitization (Figures ~\ref{fig:dvws1}, ~\ref{fig:dvws2}, and ~\ref{fig:dvws3}).

This variability underscores a key challenge: Copilot’s output is non-deterministic. Identical or near-identical code may result in different levels of scrutiny across review passes. In a security context, such inconsistency undermines trust and reproducibility. Moreover, even when security-relevant comments were generated, they did not explicitly name the vulnerabilities, lacked severity ratings, and did not cite formal sources (e.g., OWASP or CWE IDs).
%\subsubsubsection{Appendix F: DVWS Node Test Details}
%\label{appendix:dvws}
%On Re-Review Request, Copilot reviewed 46 out of 50 changed files in this pull request and generated 7 comments.

%Summary of discoveries (bolded has the screenshot attached): broken html tag, improvement on header construction, check for changes on a field before re-calculation, properly identify password field, variable declaration improvement, improper sanitization, spelling mistake.

\subsection{Failure to Interpret Context and Propagation}
A recurring weakness across all evaluations was Copilot’s inability to reason about data flow across functions or files. Security vulnerabilities often involve interprocedural flows—e.g., user input being passed through multiple layers before reaching a sink (e.g., database query). Copilot’s model appears limited to shallow, token-based reasoning within a narrow context window, meaning it cannot trace or correlate user input to exploitable sinks.

This contrasts sharply with CodeQL, which builds a program’s semantic graph and can perform complex path queries. For example, a CodeQL query can detect SQL injection patterns involving multiple functions and even dynamic query assembly. Copilot, by contrast, operates more like a statistical pattern matcher, unaware of control/data dependencies.

\subsection{Coverage Limitations and Language Support}
While Copilot supports a broad range of languages (as shown in Table~\ref{tab:table1}), our results suggest it struggles with non-mainstream file types and configuration files. For example, YAML, HTML, XML, and ASP files were either ignored or incompletely reviewed. Since many security vulnerabilities originate in configuration and integration code—such as misconfigured CORS headers or API keys in environment files—this gap represents a critical limitation for enterprise use.

\begin{table*}[t!]
\centering
\renewcommand{\arraystretch}{1.2}
\begin{tabularx}{\textwidth}{|p{2.8cm}|p{3cm}|X|p{4cm}|}
\hline
\textbf{Dataset} & \textbf{Category / Languages} & \textbf{Known Vulnerabilities} & \textbf{Copilot Code Review Results} \\
\hline
Allsafe \newline \url{https://github.com/t0thkr1s/allsafe} & Mobile Apps \newline Java (59.6\%), Kotlin (36.4\%), C++ (2.7\%) & 
Insecure Logging, Hardcoded Credentials, Root Detection, Arbitrary Code Execution, Secure Flag Bypass, Certificate Pinning Bypass, Insecure Broadcast Receiver, Deep Link Exploitation, SQL Injection, Vulnerable WebView.
Labeled by: Authors. & 
117/123 files reviewed, 4 comments generated. AndroidManifest.xml not reviewed. Detected NullPointerException, RuntimeException.(\ref{appendix:allsafe}) \\
\hline
WebGoat \newline \url{https://github.com/jerryhoff/WebGoat.NET/} & Web App \newline C\# (72.6\%), Classic ASP (16.8\%), CSS (7.1\%), Javascript (3.5\%) &
SQL Injection, Broken Authentication, Sensitive Data Exposure, XML External Entities, Broken Access Control, Cross-Site Scripting, Insecure Deserialization, Vulnerable Components, Request Forgeries.
Labeled by: OWASP. & 
1011/1019 files reviewed, 1 comment (typo). 100\% Java subset: 174/174 files reviewed, 1 typo comment. On re-review: Hardcoded indices discovered (\ref{appendix:webgoat}). \\
\hline
SARD XSS Test Case 501377 \newline \url{https://samate.nist.gov/SARD/test-cases/501377/versions/1.0.1} & Web Apps \newline Java, ASP, JSP &
CWE-79 Cross-Site Scripting vulnerability. Labeled by: SAMATE Team. &
6/9 files reviewed, no comments. (\ref{appendix:cwe79}) \\
\hline
SARD SQL Injection Test Case 156515 \newline \url{https://samate.nist.gov/SARD/test-cases/156515/versions/1.0.0} & Web Apps \newline Java, YAML &
CWE-564 SQL Injection via Hibernate. Labeled by: SAMATE Team. &
4/9 files reviewed, 2 comments (spelling mistakes). (\ref{appendix:cwe79}) \\
\hline
IARPA STONESOUP Phase 3 Wireshark 1.8.0 \newline \url{https://samate.nist.gov/SARD/test-suites/22} & Desktop Apps \newline C &
127 CSVs covering multiple CWEs including CWE-457 (Uninitialized Variable), CWE-789 (Memory Allocation), CWE-119 (Memory Buffer), CWE-170 (Null Termination), CWE-835 (Infinite Loop), and CWE-670 (Incorrect Control Flow). &
878/898 files reviewed, no comments. (\ref{appendix:wireshark}) \\
\hline
SARD Sensitive Cookie Test Case 141874 \newline \url{https://samate.nist.gov/SARD/test-cases/141874/versions/1.0.0} & Web Apps \newline Java &
CWE-614 Sensitive Cookie Without Secure Attribute vulnerability. &
7/11 files reviewed, 2 comments (spelling mistakes). (\ref{appendix:cookie}) \\
\hline
dvws-node \newline \url{https://github.com/snoopysecurity/dvws-node} & Web Service and API \newline JavaScript (58.7\%), HTML (40.1\%), Other (1.2\%) &
Vulnerabilities including Insecure Direct Object Reference, Access Control Issues, Mass Assignment, Cross-Site Scripting, NoSQL Injection, Server-Side Request Forgery, JWT Secret Key Brute Force, CORS Misconfigurations, SQL Injection, XXE, Command Injection, Open Redirect, GraphQL vulnerabilities, among others. &
46/50 files reviewed initially, 0 comments. Upon re-review, 7 comments made including proper password field identification and sanitization improvements. (\ref{appendix:dvws}) \\
\hline
\end{tabularx}
\caption{Summary of Copilot Code Review Performance Across Different Datasets}
\label{tab:evaluation}
\end{table*}
\subsection{Consistency, Explainability, and Practicality}
Even in cases where Copilot did flag potentially problematic code (e.g., field naming in \texttt{dvws-node}), the feedback lacked structure, severity indicators, or links to remediation guidance. Traditional tools like CodeQL or SonarQube offer line-precise diagnostics, CWE tagging, and fix suggestions. Copilot’s vague comments (“you might want to rename this variable”) offer little value in triage workflows or secure CI/CD pipelines.

Additionally, the inconsistency in review behavior across re-submissions raises concerns about reproducibility. Developers relying on AI reviews need consistent, explainable results, especially in regulated industries (e.g., healthcare, finance), where traceability and auditing are mandatory.

\subsection{Summary and Key Takeaways}
Our empirical results lead to the following key observations:
\begin{itemize}
    \item Copilot Code Review is currently ineffective for detecting a wide range of known security vulnerabilities.
    \item Comments generated were infrequent, inconsistent, and largely unrelated to known CWEs—even in ideal test conditions.
    \item Copilot's review model lacks semantic understanding of execution flow, data propagation, and contextual policy enforcement.
    \item When feedback was given, it was rarely actionable, lacked prioritization, and was expressed in vague or generic language.
    \item Non-deterministic behavior across review passes suggests instability and a lack of formal guarantees.
\end{itemize}

These findings suggest that GitHub Copilot, in its current review incarnation, is not ready to replace or even supplement dedicated security analysis tools. While it may be a useful assistant for catching stylistic issues or refactoring suggestions, its application in secure software development requires significant refinement, targeted training, and validation.
%\subsection{Appendix G: Ask Copilot Feature}
%\label{appendix:chat}
%Copilot Chat feature (Ask Copilot---GPT-4o).
%While Copilot Code Review feature was not able to review AndroidManifest.xml file as shown in the image below, \textbf{Ask Copilot} feature is able %to analyze and detect vulnerabilities in the same file when prompted to code-review it.
The summarized results, including project domains, programming languages, vulnerabilities present, and Copilot’s performance, are presented in Table~\ref{tab:evaluation}.

%The details and screenshots of the experiments carried in this project are includes in the attached appendices. 
\section{Conclusion}
Despite the impressive advancements of GitHub Copilot Chat in understanding and analyzing complex codebases, the evaluation of the new Copilot Code Review feature revealed surprisingly limited effectiveness in detecting known security vulnerabilities. Across a range of datasets intentionally populated with critical flaws---including SQL injection, cross-site scripting, insecure deserialization, and memory corruption---Copilot Code Review often failed to identify or comment on these issues. While it reliably generated comments on minor concerns such as typos or coding best practices, it almost never directly flagged the serious security flaws deliberately introduced into the codebases. In some cases, it only surfaced indirect indicators, such as null pointer exceptions or runtime errors, which could loosely correlate with insecure programming but did not explicitly highlight the underlying vulnerability. This contrast between Copilot Chat’s sophisticated, conversational analysis abilities and the comparatively superficial behavior of Copilot Code Review suggests that, while promising, the automated review system is not yet a reliable substitute for expert manual security auditing.
\bibliographystyle{plain}
\balance

\end{document}